\documentclass[10pt]{article}
\usepackage[english]{babel}
\usepackage{fullpage}
\usepackage{url} 
\usepackage[dvipsnames]{xcolor}
\usepackage{listings}
\usepackage{tabularx} 
\usepackage{amsmath}  
\usepackage{graphicx} 
\usepackage{amsfonts}
\usepackage{booktabs}
\usepackage{bbm}
\usepackage{xcolor}
\usepackage{scalerel,stackengine}
\usepackage{xspace}
\usepackage{soul}
\usepackage[textwidth=2cm]{todonotes}

\usepackage{pdfpages}

\makeatletter
\if@todonotes@disabled

\else

\fi
\makeatother

\definecolor{Gold}{rgb}{1, 0.84, 0}
\definecolor{Green}{rgb}{0.0, 0.5, 0.0}
\definecolor{Indigo}{rgb}{0.29, 0.0, 0.51}
\definecolor{Crimson}{rgb}{0.86, 0.08, 0.24}

\DeclareMathOperator*{\argmin}{arg\,min}

\newcommand{\harpa}{\textsc{Harpa}\xspace}

\definecolor{revgreen}{RGB}{20,150,20}

\title{High-Rate Phase Association with Travel Time Neural Fields}

\author {Cheng Shi$^{1}$, Giulio Poggiali$^{2}$, Chris Marone$^{2,5}$, Maarten V. de Hoop$^{3,*}$ and Ivan Dokmanić$^{1,4,*}$\\
\normalsize{$^{1}$Departement of Mathematics and Computer Science, University of Basel,}\\
\normalsize{$^{2}$Department of Earth Sciences, La Sapienza Università di Roma,}\\
\normalsize{$^{3}$Simons Chair in Computational and  Applied  Mathematics and Earth Science, Rice University,}\\
\normalsize{$^{4}$Department of ECE, University of Illinois at Urbana--Champaign,}\\
\normalsize{$^{5}$Department of Geosciences, Pennsylvania
State University}\\
\normalsize{$^\ast$To whom correspondence should be addressed; E-mail:  ivan.dokmanic@unibas.ch, mvd2@rice.edu}
}
\begin{document}
\maketitle

\begin{abstract}
Earthquake science and seismology rely on the ability to associate seismic waves with their originating earthquakes. Earthquake detection algorithms based on deep learning have progressed rapidly and now routinely detect microearthquakes with unprecedented clarity, providing information about fault dynamics on increasingly finer spatiotemporal scales.  However, this densification of detections can overwhelm existing techniques for phase association which rely on fixed wave speed models and associate events one by one. These methods fail when the event rates become high or where the 4D complexity of elastic wave speeds cannot be ignored. Here, we introduce \harpa, a deep learning solution to this problem. \harpa is a high-rate association framework which incorporates wave physics by leveraging deep generative models and travel time neural fields. Instead of associating events one by one, it lifts arrival sequences to probability distributions and compares them using an optimal transport metric. The generative travel time neural fields are used to estimate the wave speed simultaneously with association. \harpa outperforms state-of-the-art association methods for both real seismic data and complex synthetic models and paves the way for improved understanding of seismicity while establishing a new seismic data analysis paradigm.
\end{abstract}

\section{Introduction}

Seismic signals are a gateway to understanding the structure and dynamics of the Earth and the physics of earthquakes and seismic faulting. Recent breakthroughs in deep-learning-based seismic processing enable reliable detection of small magnitude events~\cite{ross2018generalized, mousavi2020earthquake, zhu2019phasenet}, with ongoing developments promising to reduce  detection thresholds even further ~\cite{liu2024seislm}.
This has the potential to reveal unprecedented details about the variability of elastic material properties in Earth's crust and upper mantle and shed light on nonlinear earthquake dynamics at increasingly finer scales. But progress is hindered by the limitations of existing techniques for identifying and grouping seismic wave arrivals from their originating earthquakes---the so-called seismic phase association problem.
These techniques have been primarily designed for lower-rate, larger magnitude events. But in accord with the Gutenberg--Richter law, as magnitude detection thresholds decrease, the corresponding high arrival rates pose unique combinatorial challenges to association~\cite{grechka2017microseismic}.  

\begin{figure}[t!]
  \includegraphics[width=1\textwidth]{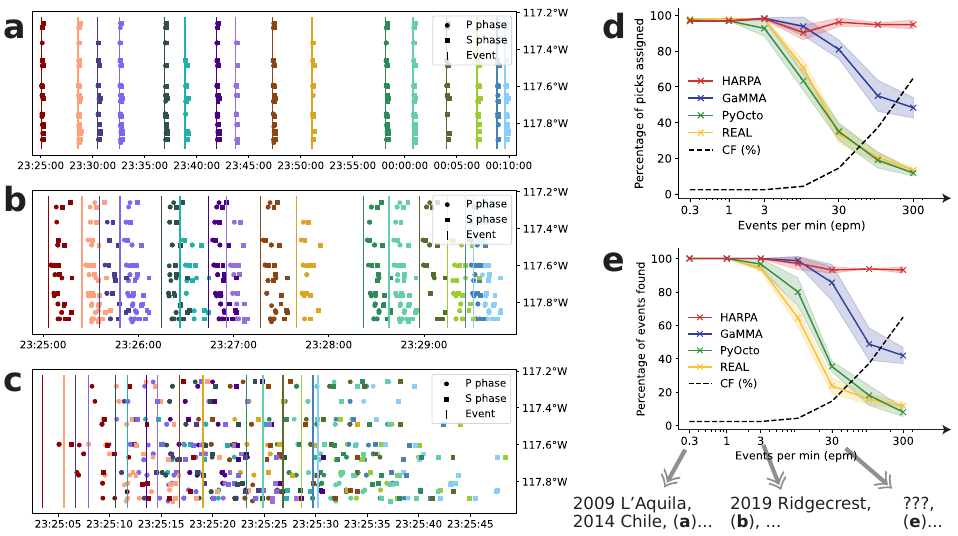}
  \caption{Phase arrivals at different event frequencies. Each row of Panels a-c shows data for a given seismometer location (latitude at right) as a function of time. Event rates are (\textbf{a}) 0.3,  (\textbf{b}) 3, and (\textbf{c}) 30 per minute. Panel (\textbf{b}) shows real seismic data from the 2019 Ridgecrest earthquake, while (\textbf{a}) and (\textbf{c}) are generated by shifting arrivals and events from (\textbf{b}) to produce a range of event rates for phase association testing. (\textbf{d}) and (\textbf{e}) show performance comparison of \harpa~with other algorithms across different frequencies of events. Note that \harpa outperforms all associators for high event rates. $\mathsf{CF}$ is the \emph{confusion factor} representing the degree of arrival complexity based on phase overlap and ranges from 0 (easy) to 1 (difficult) as detailed in the \textbf{SI}.}\label{fig: semi}
\end{figure}

Fig.~\ref{fig: semi} illustrates the phase association problem using data from two recent earthquake sequences where the phases arrive at the stations in a scrambled, complex order. The combination of unknown, seemingly random permutations and high arrival rates renders algorithms based on grid search ineffective. Moreover, for geologically complex areas where the seismic wave speed is highly inhomogeneous~\cite{sato2012seismic}, the usual constant or depth-only dependent approximations become a limiting factor for seismic phase association. Errors caused by overly simplistic seismic wave speed models are minimal when earthquake rates are low but pose a critical challenge for high-rate events. Although existing algorithms can be adapted to incorporate more complex wave speed models for travel time computation~\cite{ross2023neural}, the problem is that the true models are rarely known at the required level of accuracy: the challenge lies in adaptability rather than parametrization.

To address these issues, we propose \harpa---a framework for high-rate seismic phase association with either known or unknown wave speed. \harpa is effective in settings that are much more complex than those addressed by existing methods. It first estimates locations and occurrence times of earthquakes jointly with an approximate wave speed model and then uses the estimated parameters to assign earthquake phases by solving a much simpler linear assignment problem. This simplification is afforded by leveraging the physics of wave propagation and deep learning advances in generative modeling, optimal transport, and stochastic optimization.

\begin{figure}[!t]
    \centering
    \includegraphics[width=1\textwidth]{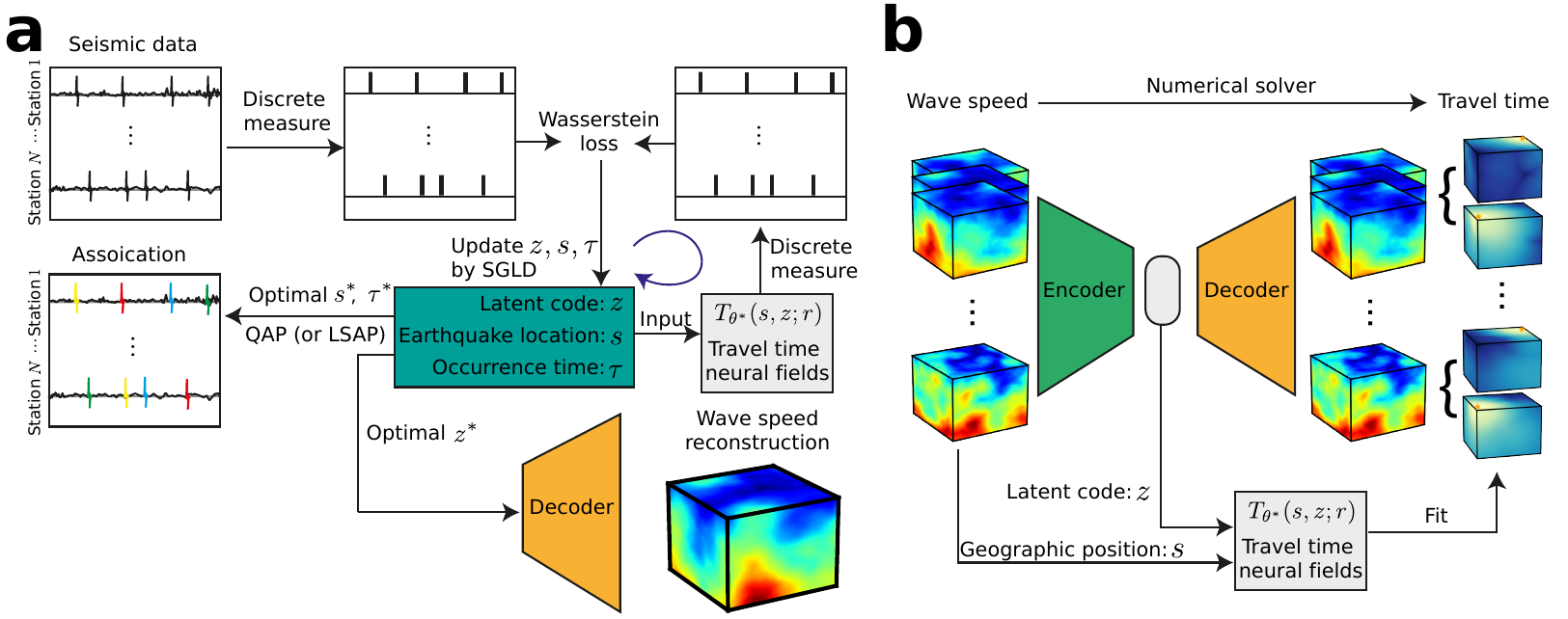}
    \caption{Workflow for travel time neural field implementation with optimal transport. (\textbf{a}) Diagram of \harpa's association pipeline. Note connections between discrete measurements, latent code and optimal association. (\textbf{b}) Diagram of the wave speed autoencoder and travel time neural field with encoder/decoder stages. QAP, LSAP, and SGLD refer to the Quadratic Assignment Problem, the Linear Sum Assignment Problem, and Stochastic Gradient Langevin Dynamics, respectively.
    }
    \label{fig: diagram}
\end{figure}

To address the first problem---highly entangled arrivals (Fig.~\ref{fig: semi}\textbf{c})---\harpa~interprets the arrival time data at each station as a probability distribution and uses an optimal transport metric to quantify the discrepancy between the distributions of observed and estimated data. This workflow (Fig.~\ref{fig: diagram}) obviates the traditional one-to-one matching between phases and events, replacing it by matching between distributions of arbitrary numbers of arrivals, and leads to a simultaneous estimation of all event locations and occurrence times.  Fig. SF3 of the \textbf{SI} shows seismograms corresponding to these cases.

To address the second problem---unknown wave speed---we first train a deep generative model to embed a family of complex wave speeds in a low-dimensional latent space (Fig.~\ref{fig: diagram}). We then use this generative model in \harpa to simultaneously estimate the wave speed model and the spatio-temporal earthquake parameters. This is illustrated in Fig.~\ref{fig: diagram}.  Instead of directly addressing the formidable challenge of optimizing the source parameters through a numerical solver we train a deep neural field (also known as implicit neural representation)~\cite{sitzmann2020implicit, mildenhall2021nerf, xie2022neural} fitting the travel times between arbitrary interior (source) points and a fixed set of receivers \textit{for an entire family of wave speeds}. An autoencoder structures the latent space and provides a pairing between latent codes and wave speeds that is then used by the generative travel-time neural field. As we show in Section~\ref{sec: Experiments} and in \textbf{Supplementary information (SI)}, fitting the best wave speed from the range of a sufficiently diverse generative model makes the association highly robust, even when the test wave speed is very different from those used to train the model.

We jointly determine the earthquake source parameters and the wave speed parameters by differentiable optimization of the discrepancy between observed and model probability distributions of arrivals. While the problem of wave speed reconstruction from unknown sources admits a unique solution under certain conditions~\cite{de2021stable}, identifying this solution by optimizing a non-convex objective is complicated by poor local minima. Grid search algorithms could in principle resolve this, and indeed provide an effective solution in the low-rate regime with simple wave speeds. But they become extremely slow and inaccurate in the high-rate regime~\cite{munchmeyer2024pyocto} where the loss landscape is complex.

We thus include the third key ingredient for phase association in complex, high event-rate cases: we optimize the model parameters using stochastic gradient Langevin dynamics (SGLD)~\cite{welling2011bayesian}, a combination of stochastic gradient-based methods and sampling. Markov Chain Monte Carlo (MCMC) methods can in principle deal with the non-convexity but they are ineffective in our setting because the search space is too large. SGLD effectively addresses this problem while ensuring an adequate exploration of the loss landscape, ultimately leading to the identification of a (near-)global optimum. 

\begin{figure}[!t]
    \includegraphics[width=1\textwidth]{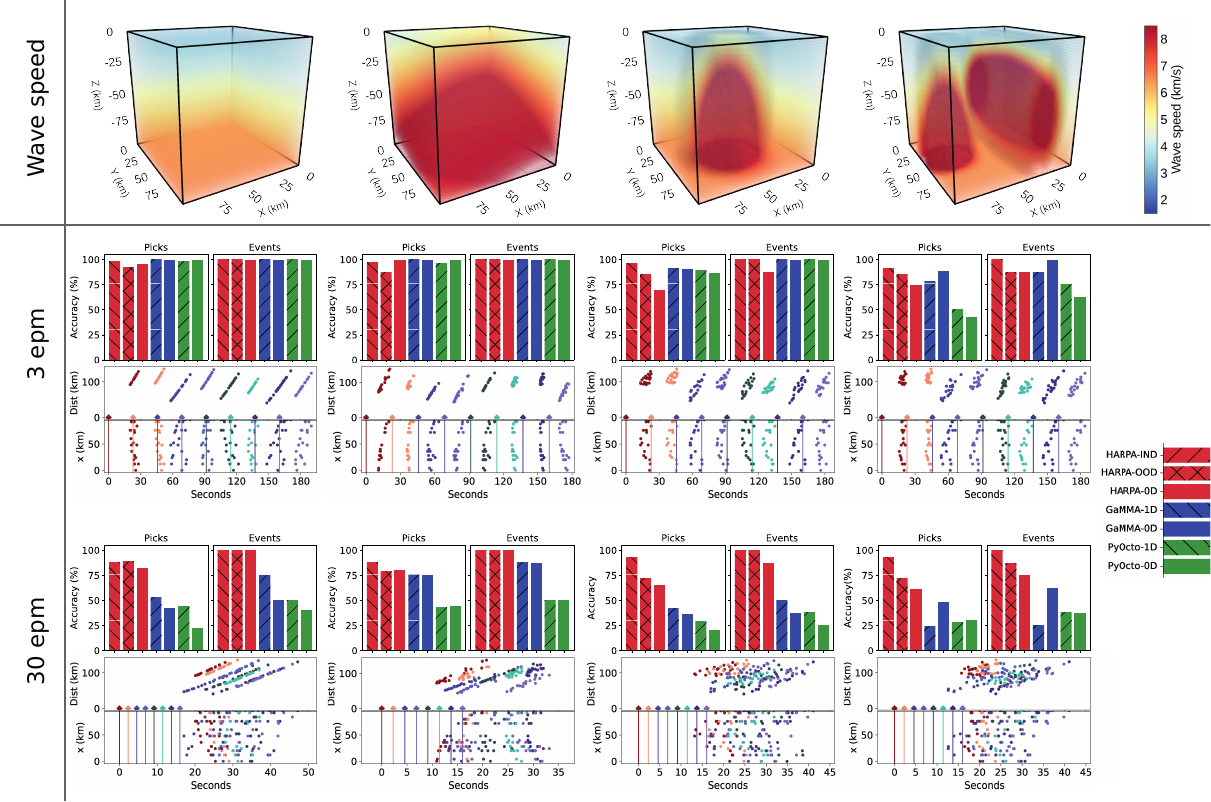}
    \caption{Association for four wave speed models (top) and event frequencies of 3 and 30 per minute. The histograms illustrate the accuracy of various algorithms, with \textit{0D} and \textit{1D} denoting constant and one-dimensional wave speed models, respectively. \textit{IND} and \textit{OOD} represent in-distribution and out-of-distribution wave speed models used for testing against the trained generative model for \harpa. In the scatter plots, the top row shows the relationship between station-event distance and arrival time, while the bottom row presents x-coordinates versus arrival time. Note that \harpa associates with high accuracy even when event rates are high in areas with complex wave speed models.}\label{fig: wave speed}
\end{figure}

\subsection*{Seismic Phase Association}

Over the last decade, there has been increasing interest in the application of machine learning in seismology, for example, in seismic denoising~\cite{trappolini2024cold}, phase picking~\cite{ross2018generalized,mousavi2020earthquake,zhu2019phasenet,zhu2019deep}, travel-time computation~\cite{smith2020eikonet,anderson2023emulation}, and wave speed ``inversion''~\cite{gao2021deepgem}. Machine learning approaches to phase association use either probabilistic ~\cite{ross2019phaselink,zhu2022earthquake,ross2023neural} or graph models~\cite{mcbrearty2019earthquake,mcbrearty2023earthquake}. Some algorithms use fast phase grid search or grid division approaches~\cite{zhang2019rapid,munchmeyer2024pyocto}. However, when arrivals are highly scrambled, these approaches become either exceedingly slow or inaccurate. Some of the methods adopt a homogeneous or a depth-only dependent wave speed model~\cite{zhu2022earthquake}, while others incorporate wave speed information indirectly through supervised learning. In the latter case, the association on a specific wave speed can be learned by fitting the association labels of synthetic events whose arrivals are generated based on the same (approximate) wave speed model~\cite{ross2019phaselink,mcbrearty2023earthquake}. Since supervised approaches do not explicitly leverage wave physics they demand large quantities of data for training. Implicit representations have been recently used for association with more complex but fixed wave speeds models~\cite{smith2020eikonet, ross2023neural}.

But the fundamental challenge is that the wave speed model is not known, certainly not at the resolution required to effectively process data from high-rate seismicity. Furthermore, the event-by-event association paradigm common to all existing methods becomes ineffective at higher seismicity rates even when the wave speed is fixed and simple. We show that rethinking the association problem in terms of probability distributions of arrivals enables a more effective, implicit utilization of wave propagation models from the outset.

There are also recent studies that directly locate events from continuous waveforms, thereby bypassing the phase association problem~\cite{shi2022malmi,liu2024dmloc,isken2024qseek}. These methods, based on so-called migration in wave speed models---which are assumed known---identify events by searching for peaks in stacked back-projections. Although this avoids dealing with disordered arrivals, it is complicated by waveform inaccuracies and heterogeneous magnitudes, particularly when arrivals become dense. As a result, these algorithms are applied to settings with tens of events per day. Our work rather aims at operating with tens of events per minute (epm).

\paragraph{Connections with combinatorial optimization and inverse problems} 

If earthquake source locations, occurrence times, and wave speeds were all known, the phase association problem could be formulated as a linear sum assignment problem (LSAP), which is effectively solved by the Hungarian algorithm~\cite{burkard2012assignment}. If only the occurrence times are unknown, the problem is related to a quadratic assignment problem (QAP)~\cite{koopmans1957assignment, pardalos1994quadratic} which is much harder to solve (it belongs to the class of NP-hard problems). While QAPs can be solved in certain cases using heuristics and meta-heuristics~\cite{ahuja2000greedy,benlic2013breakout}, there has been growing interest in applying machine learning to hard combinatorial problems~\cite{vinyals2015pointer, li2018combinatorial, nowak2018revised}. The situation, however, is often even more complex: neither the wave speed nor the spatiotemporal earthquake source locations are known, although sometimes a rather coarse wave speed model is available. These unknown degrees of freedom result in a mixed-integer programming problem which is harder than LSAP or QAP and challenging to solve even with a small number of earthquakes and stations~\cite{floudas1995nonlinear, geist2018determining, belotti2013mixed}. 

The joint localization--wave speed recovery problem has a unique solution under certain conditions~\cite{de2021stable}, but there are no reliable algorithms that solve it provably and efficiently. There exist algorithms that attempt joint full waveform (or sometimes travel time) inversion and wave speed model refinement but with temporally well-separated sources, thus implicitly avoiding the association problem (but not the ill-posedness)~\cite{sun2016full,witten2017image,song2019microseismic}. In the context of micro-seismicity, simple parametric models~\cite{jansky2010feasibility} and Bayesian priors~\cite{zhang2017simultaneous} were used for regularization. In our case the prior implicit in the generative model ensures a stable reconstruction. We mention that our strategy to convert combinatorial data into measures in order to avoid combinatorial optimization is reminiscent of earlier work on unassigned Euclidean distance data~\cite{dokmanic2013acoustic,huang2021reconstructing} with applications in genomics and room acoustics.

\section{Materials and Methods}\label{sec: Materials and Methods}

We consider an unknown wave speed $c$ on some spatial domain of interest $\Omega$ which contains all sources and ray paths of seismic waves. We assume that $c$ belongs to a family of plausible wave speed models $\mathcal{C}$. For a given $c$ we denote by $T_c(s;r)$ the time it takes for a wave to travel from a source $s\in \Omega$ to a receiver $r \in \partial \Omega$ (with $\partial \Omega$ being boundary of $\Omega$). We assume that there are $N$ stations, $M$ sources, and each station receives $K$ phases originating from these $M$ sources. For simplicity we will first assume $K = M$ and later discuss situations with noisy and missing picks where $K \neq M$. We assume that earthquakes are well-approximated by a volumetric distribution of point sources that generate sufficiently rich data, which includes complex cases like swarms where events occur on faults at varying orientations or in ``clouds''~\cite{zang2014analysis,glubokovskikh2023transforming}.

We consider a set of stations with locations $\mathcal{R}=\left\{r_1, \ldots, r_N\right\}$. For a given time window, we are interested in earthquakes with spatio-temporal locations $\mathcal{E} = \{(s_j,\tau_j)\}_{j=1}^M$, where $s_j \in \Omega$ is the location and $\tau_j$ the occurrence time of event $j$. Arrival times at the $i$th station are collected in the set $\mathcal{D}_i=\{t_{i,j}\}$, where $t_{i,j}=T_c(s_j;r_i)+\tau_j$ is the arrival time for each pick. Instead of considering the arrivals individually as in existing approaches, we model the \emph{probability distribution} of arrival times within a window. For the observed arrivals at station $i$ this is a discrete probability distribution concentrated on arrival times,
\begin{equation}
    \mathbb{P}_{\text{obs}, i} =  \frac{1}{M} \sum_{j=1}^M \delta_{t_{i,j}}.
\end{equation}
$\mathbb{P}_{\text{obs}, i}$ depends on the wave speed $c$ and on the spatio-temporal locations of events $\mathcal{E}$. If we can identify $c$ and $\mathcal{E}$, the association problem is solved, but this requires first solving a seemingly harder joint recovery problem. 

Our strategy will be to search for $c$ and $\mathcal{E}$ which minimize the distance between the observed probability distribution $\mathbb{P}_{\text{obs}, i}$ and the model probability distribution over arrivals $\mathbb{Q}_{(c, \mathcal{E}), i}$ induced by $c$ and $\mathcal{E}$, 
\begin{equation}\label{eqn: final optimization}
    \begin{aligned}
        \underset{\mathcal{E},c_S, c_P}{\mathrm{minimize}}& \quad \mathcal{L}(\mathcal{E},c_S,c_P) \quad
        \text{where} \quad \mathcal{L}(\mathcal{E},c_S,c_P)=\sum_{i=1}^N \text{dist}\left(\mathbb{Q}^\text{S}_{(c_\text{S}, \mathcal{E}), i},\mathbb{P}_{\text{obs}, i}^\text{S}\right)+\text{dist}\left(\mathbb{Q}_{(c_\text{P}, \mathcal{E}), i}^\text{P},\mathbb{P}^\text{P}_{\text{obs}, i}\right),
    \end{aligned}
\end{equation}
where we model both the P- and S- wave speeds, indicated in sub/superscripts. We use the optimal-transport-theorethic Wasserstein-2 distance (sometimes referred to as the earthmover's distance in the discrete 1D case) to measure the discrepancy between the model and observed arrival distributions (see \textbf{SI}). It measures the cost of optimally transporting one distribution to the other and can be computed efficiently in our case of discrete distribution of scalar arrival times.

We will show that by recasting the problem in terms of distributions, we obviate the difficulties of the usual event-by-event association. If arrival times are accurately recorded, the optimal solution achieves a near-zero loss $\mathcal{L}$ at each station $i$. After obtaining $\mathcal{E}$ from \eqref{eqn: final optimization}, we compute an association by solving a simple linear assignment problem. A detailed description of association and the corresponding mixed-integer optimization problem can be found in the \textbf{SI}.

For simplicity we assumed that there are no spurious and missing travel-time picks: Each station observes exactly $M$ arrivals. In reality quite many events are not recorded by every station, at least not at the SNR required by phase picking algorithms. The latter will in turn miss some arrivals, or, due to noise or finiteness of the domain of interest, output spurious arrivals. These effects can significantly impact association accuracy. To address this problem within our workflow, we use an unbalanced optimal transport mismatch between arrival sets which allows for different numbers of picks per station~\cite{liu2022sparsity}; the details are laid out in the \textbf{SI}.

\subsection*{Generative modeling of low-dimensional wave speed families}

To build \harpa~we assume that the wave speed models and consequently travel times exhibit a latent low-dimensional structure. We thus model the wave speeds by an autoencoder-based generative model and travel times by matching a continuous coordinate-based neural field~\cite{sitzmann2020implicit, mildenhall2021nerf, xie2022neural}. The assumption of latent low-dimensionality is consistent with geodynamical models which generate the distribution of the elastic material properties in the interior of Earth. Related ideas have been explored earlier to model Earth's mantle~\cite{shahnas2018inverse,mora2023models}. In a similar vein, geological random fields have been used to in conjuction with accurate earthquake simulations to study the 2019 Le Teil earthquake in a region where little geophysical data is available~\cite{lehmann2022machine}. The wave speeds generated by our latent model are considerably more complex than the fixed models used for association in the recent literature which are either constant or only vary along the depth coordinate~\cite{liu2020rapid, zhu2022earthquake, ross2023neural}.  

We thus define the associated travel time function as $T(s, z;r) = T_{c(z)}(s; r)$, where $z$ is the low-dimensional latent code corresponding to wave speed $c(z)$. The latent code can be shared across \textit{P}- and \textit{S} wave speeds when both kinds of phases are available. We then rewrite the optimization problem \eqref{eqn: final optimization} over $(c_P, c_S)$
as one over the latent code $z$,
\begin{equation}\label{eqn: minimize unknown z e}
    \underset{\mathcal{E},c_S\in \mathcal{C}_S, c_P\in\mathcal{C}_P}{\mathrm{minimize}} \quad \mathcal{L}(\mathcal{E},c_S,c_P) \longrightarrow \underset{\mathcal{E},z}{\mathrm{minimize}} \quad { \mathcal{L}(\mathcal{E}, c_S(z), c_P(z))}.
\end{equation}

\subsection*{Generative travel time neural fields}

The complexity of iteratively solving~\eqref{eqn: minimize unknown z e} is dominated by the complexity of computing the travel time function $T(s, z; r)$. One option is to use a standard numerical solver to compute $T$ at a desired set of points. But this poses serious challenges: numerical solvers work on grids or meshes that are difficult to interface with our continuous optimization for source locations. This is further complicated by the need to use differentiable optimization: in order to jointly optimize over sources and wave speeds we must differentiate $T$ with respect to both the source location and the latent code $z$. Regardless of the method used to represent wave speed models, this would result in exceedingly expensive computations.

We resolve this conundrum in \harpa by first using a deep autoencoder to build a low-dimensional latent representation of a set of complex wave speeds (Fig.~\ref{fig: diagram}) and then fitting a neural field~\cite{mildenhall2021nerf,sitzmann2020implicit} to numerically compute travel times between a fixed set of receivers and sources at each possible mesh node. We use the fast marching method as implemented in \texttt{scikit-fmm}~\cite{scikit-fmm} to obtain ground-truth travel times for training. We use a neural network $T_\theta$ with weights $\theta$ and periodic activation functions~\cite{sitzmann2020implicit}, which ensures that travel times and their derivatives with respect to source locations and wave speed latent codes can be efficiently evaluated
at continuous coordinates. This is a generalization of recent neural models of travel times  for a \textit{fixed} wave speed \cite{smith2020eikonet, ross2023neural}. Instead of training neural fields to satisfy the eikonal equation~\cite{smith2020eikonet}, we simply directly train $T_\theta$ so that $T_{\theta^*}(s, z; r) \approx T(s, z; r)$ fits the travel time at a discrete grid of points, 
\begin{equation}
    \theta^*=\argmin_{\theta} \sum_{z \in \mathcal{Z}, s \in \mathcal{S}, r\in \mathcal{R}} \|T(s,z;r)-T_\theta(s,z;r)\|^2,
\end{equation}
where $\mathcal{Z}$ is a sampling of wave speed latent codes, $\mathcal{S}$ the set of grid coordinates, and $T(s, z; r)$ is pre-computed using a numerical solver. The latent travel time neural field, which mimics the real travel time function, provides a continuous representation of travel times which is differentiable with respect to both $s$ and $z$, thus enabling gradient-based optimization. 

\subsection*{Stochastic gradient Langevin dynamics}

The optimal transport distance between arrival distributions in \eqref{eqn: final optimization} is differentiable with respect to both $z$ and $\mathcal{E}$ but it is not convex in these parameters. This means that local gradient-based optimization may converge to local minima. In fact, in our experiments, (stochastic) gradient descent very often converges to \textit{poor} local minima (an example is given in the \textbf{SI}).

We resolve this by adapting stochastic gradient Langevin dynamics (SGLD)~\cite{welling2011bayesian}, an iterative optimizer that combines the favorable aspects of sampling methods like MCMC. SGLD which can be viewed either as a mini-batch variant of Langevin dynamics or as SGD with noise. The global convergence of Langevin dynamics can be leveraged to locate the globally-optimal earthquake and wave speed parameters which approximate their corresponding ground truth values, while benefiting from fast convergence rates~\cite{ma2019sampling,xu2018global}. An SGLD iterate is computed as 
\begin{equation}\label{eqn: SGLD}
       x \leftarrow x-\eta  \widehat{ \nabla_x \mathcal{L}}+\sqrt{2 \eta} \epsilon 
        \xi_x,  \quad  x \in \{z, s_1, \tau_1, \ldots, s_M, \tau_M \}
\end{equation}
where $\eta$ is the step size, and $\widehat{\nabla \mathcal{L}}$ is a mini-batch sum in \eqref{eqn: minimize unknown z e} which is an unbiased estimator of the gradient $\nabla \mathcal{L}$. The parameter $\epsilon$ modulates the amplitude of noise; $\xi_z$, $\xi_s$, and $\xi_\tau$ are standard normal vectors independently drawn across different iterations from a Gaussian distribution with mean zero and variance one.

\section{Results}\label{sec: Experiments}

\harpa performs well in comparisons with SOTA phase association models. For these tests, we used the same preprocessing for \harpa as in GaMMA to initially cluster the picks. We first divide the picks into sub-windows using the DBSCAN algorithm~\cite{schubert2017dbscan} and then to ensure a fair comparison we adopt the same max/min time results for each pick and for each event for all algorithms (Figs.~\ref{fig: semi}, \ref{fig: wave speed}, \ref{fig: chile} and~\ref{fig: ridgecrest}). Events are excluded if not sufficiently corroborated by their nearest stations~\cite{nearest-stations}. We do not leverage amplitude information of picks, although this could be a useful next step.
Further criteria and comparisons with SOTA algorithms, including event locations and time complexity, are given in the \textbf{SI}.  

\subsection{Field data, low-to-moderate rate regime and fixed wave speed model}

\begin{figure}[!h]
\includegraphics[width=1\textwidth]{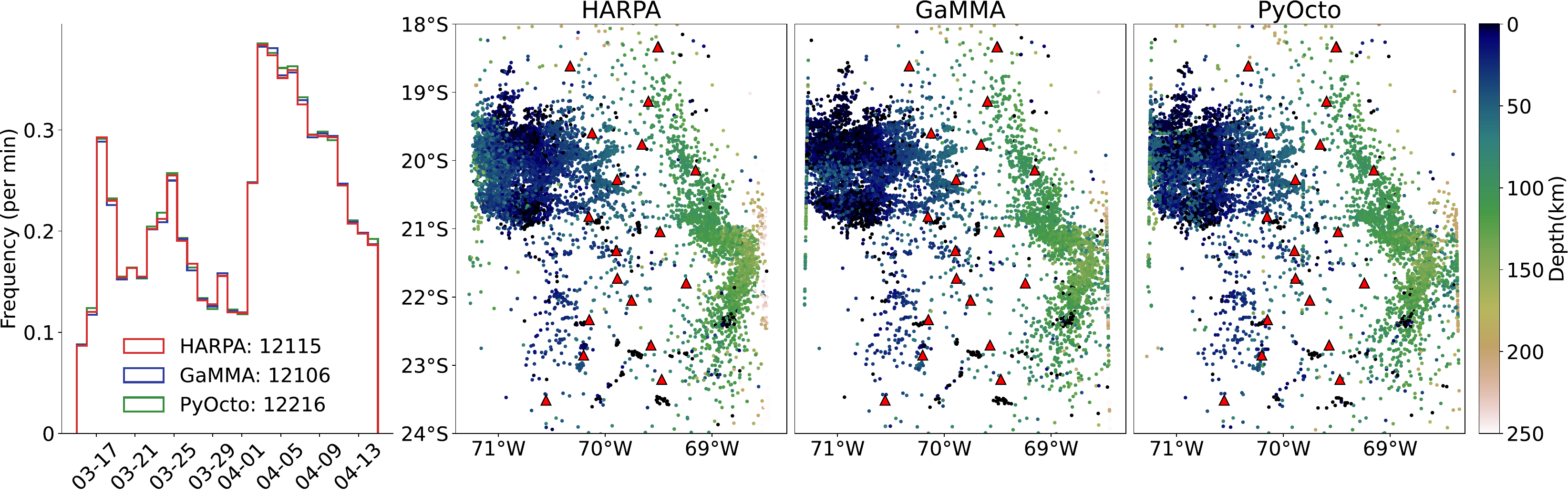}
    \caption{Performance of \harpa~and other algorithms on data for the 2014 $M_w$ 8.2 Iquique Chile earthquake. Triangles are seismic stations and dots are earthquake epicenters color coded by depth. Note that the event rates vary from about 0.1 to 0.4 per minute. For these low to moderate event rates performance is about the same for each model. }\label{fig: chile}
\end{figure}

\begin{figure}[!h]
    \includegraphics[width=1\textwidth]{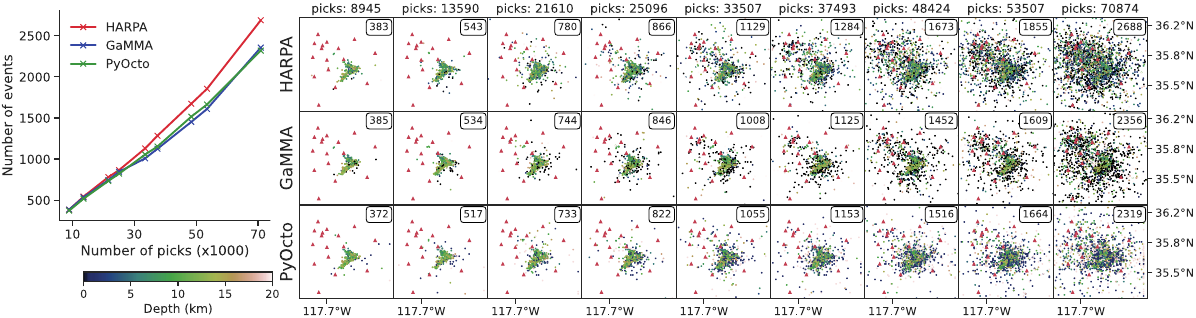}
    \caption{Performance of \harpa~and other algorithms for the 2019 $M_w$ 7.1 Ridgecrest earthquake sequence. For the map views, each row show results for one model (\harpa, GaMMA, PyOcto) and each column is a different phase pick density for the association task.  We adjusted the threshold in PhaseNet to obtain a range of event densities. The number of phase picks is given above each row of map views; number in each map is the number of events detected. Plot at left shows comparison of the events associated as a function of phase picks. \harpa performs as well or better than the other models.}\label{fig: ridgecrest}
\end{figure}

We first apply \harpa~to data from two major earthquakes: the $M_w$ 8.2 Iquique Chile earthquake of 1 April 2014, which was preceded by a $M_w$ 7.6 foreshock on 16 March 2014; and the $M_w$ 7.1 Ridgecrest earthquake of 5 July 2019, which was preceded by $M_w$ 6.4 and $M_w$ 5.4 events on 4 July 2019. For the Iquique earthquake, we use data for a one month period starting on 15 March 2014 (Fig.~\ref{fig: chile}). For the Ridgecrest sequence we use data for a 7 hour period starting at 17:00 on 4 July 2019 
(Fig.~\ref{fig: ridgecrest}). In the Chile dataset, the arrival phases (both P and S) are identified from continuous waveforms using PhaseNet~\cite{zhu2019phasenet} following the same settings as in GaMMA~\cite{zhu2022earthquake}. In this experiment the typical event rate is about 0.3 epm. Our experiments show that in this low-to-moderate rate regime, with a simple, fixed wave speed, \harpa matches the performance of the state-of-the-art algorithms GaMMA~\cite{zhu2022earthquake} and PyOcto~\cite{munchmeyer2024pyocto}. We obtain similar results on the Ridgecrest dataset with the picks obtained by PhaseNet; here the typical rate is about 3 epm (with a maximum of 7 during that week).  \harpa matches, again, the state-of-the-art performance of GaMMA and PyOcto. 

We then create a sequence of more challenging tests on Ridgecrest by progressively lowering the detection magnitude threshold in PhaseNet. As expected (see Fig.~\ref{fig: ridgecrest}) this results in an increasing number of picks, some of which are likely at or below the signal-to-noise ratio for A-quality events. \harpa nonetheless handles this case, even for events with low-signal-to-noise ratio. A formal workflow would entail a quality threshold that might eliminate lower quality detections, but here we retain all of them to create a challenging high-rate test for both \harpa and the SOTA models. For this comparison, we use PhaseNet trained within the SeisBench workflow~\cite{woollam2022seisbench} and use thresholds for both \textit{P}- and \textit{S}-phases at 0.5, 0.25, 0.05, 0.025, 0.005, 0.0025, 0.0005, 0.00025, and 0.00005, from left to right in Fig.~\ref{fig: ridgecrest}. The results show that \harpa consistently detects a higher number of sources than GaMMA and PyOcto, especially at higher rates. 
We note again that when the thresholds are set very low all association algorithms must handle a large number of potentially noisy picks. These picks can form events that pass all filtering criteria and so cannot be excluded solely at the association stage. Therefore the fact that \harpa detects a higher number of events which are consistent with all travel time information underscores its effectiveness in associating picks in the high-rate regime. 

We also emphasize that the comparisons of Figs.~\ref{fig: chile} and \ref{fig: ridgecrest} are still for a fixed, homogeneous wave speed model, which is a reasonable approximation in Ridgecrest and a setting that GaMMA and PyOcto were optimized for. We used the same  wave speed for all algorithms: 7.0 km/s for the $\mathrm{P}$ wave and 4.0 km/s for the S wave in Chile, and 6 km/s for the \textit{P} wave and 3.53 km/s for the \textit{S} wave in Ridgecrest. GaMMA and PyOcto also support a depth-only dependent wave speed model but it has a minimal effect on their performances. This is understandable in Ridgecrest where events happen at shallow depths, but it is also true in Chile, as the depth-only dependent model for P-wave velocities, ranging from 5.8 km/s at sea level to 8.6 km/s at 300 km below sea level, does not identify more events than the constant model in this regime~\cite{munchmeyer2024pyocto}.

In order to corroborate these findings we next design a robust set of experiments with field data where, unlike here, we have access to ground truth associations (cf. Fig.~\ref{fig: semi}). These experiments clearly illustrate \harpa’s advantage in dense, high-rate settings.

\subsection{High-rate experiments with densified field data}

We recall that the typical rates in Chile and Ridgecrest data are 0.3 epm (cf. Fig.~\ref{fig: semi}(\textbf{a})) and 3 epm (cf. Fig.~\ref{fig: semi}(\textbf{b})). Although in Ridgecrest arrivals begin to mildly overlap at 3 epm they remain distinguishable even by naked eye. The primary association challenge in this regime is thus to identify false and missing picks. No phase association algorithms were applied at rates higher than those reported in the 2019 Ridgecrest data in GaMMA and Neuma publications~\cite{ross2023neural}. (We focus on rate but note that factors like event depth and wave speed also modulate the problem difficulty.)

Indeed, we are not aware of existing curated benchmarks at higher rates, but both the Gutenberg--Richter law and augmented catalogs using template matching techniques (e.g. ~\cite{ross2019searching}) suggest that a wealth of small-magnitude events exist that are not included in common seismic catalogs. This is due not only to the technical difficulty of detecting such low-SNR events, but also to the density of the recording network and the quality-control requirements in the downstream seismic location pipeline. Nevertheless, this landscape is rapidly evolving. 
We anticipate that phase association and location will soon benefit from higher detection rates enabled by the latest AI-driven seismic detection tools, such as SeisLM~\cite{liu2024seislm} or ~\cite{sun2023phase}, and seismic denoising methods ~\cite{zhu2019seismic,trappolini2024cold}, coupled with an increase in the spatial density of instrumentation.

With this in mind, we designed another test by manually creating high-rate seismograms using data from the 2019 Ridgecrest sequence. To do this we extracted a time window in which ground truth events and their associated picks are available in an existing catalog, at a detection threshold sufficiently high that they can be considered ground truth (Fig.~\ref{fig: semi}). We then advanced or delayed occurrence times of events by consistently adjusting arrival times between stations. This adjustment leaves the wave travel paths intact, resulting in highly-realistic high-rate data. The same strategy can also be applied to seismograms (cf. the \textbf{SI}).

As seen in Fig.~\ref{fig: semi}(\textbf{c}), when the rate increases to 30 epm, arrivals mix significantly and association is markedly more complex. All algorithms perform similarly at rates below 3 epm. However, as the rate increases and the association problem becomes more challenging, only \harpa~remains effective. We emphasize that unlike grid-search algorithms which  become extremely (exponentially) slow in high-rate regimes with a large number of arrivals, the runtime of \harpa, which lifts any number of arrivals to a probability distribution, is minimally affected by event frequency; a detailed comparison is given in the \textbf{SI}.

\subsection{Unknown wave speed model: Experiments with synthetic data and low- to high-rate regimes}

In low-rate settings with simple wave speed models (cf. Fig.~\ref{fig: semi}(\textbf{a}) or the second row in Fig.~\ref{fig: wave speed}), the accuracy of the wave speed model has a limited impact on association performance. But in the high-rate regime (cf. Fig.~\ref{fig: semi}(\textbf{c})), a more accurate wave speed model becomes essential. To show this clearly, we designed a set of experiments with heterogeneous synthetic wave speed models along with noiseless P-wave data. The wave speed models (shown in Fig.~\ref{fig: wave speed} together with arrivals) are created by augmenting a vertical gradient wave speed (the first wave speed model in Fig.~\ref{fig: wave speed}) with smooth ellipsoidal perturbations. The top row shows four wave speed models from the training distribution. We compare \harpa's performance with GaMMA and PyOcto, with both constant and 1D models. Crucially, we also run an out-of-distribution experiment (see \textbf{SI}) where we compare an instance of \harpa trained on Gaussian random fields which are qualitatively very different distribution from wave speeds that we are applying the model to.

Comparing the results at different rates clearly shows that in the high-rate regime, neither a constant nor a 1D approximation used in existing algorithms suffices for effective association. This highlights the need for better, \textit{adaptive} wave speed modeling. We emphasize that the out-of-distribution model still significantly outperforms a constant wave speed instance of \harpa (and all baseline algorithms) in high-rate regime, which further highlights the importance of wave speed adaptivity. High-resolution differentiable parametrizations of wave speed models such as those employed by Neuma~\cite{ross2023neural} (and also here) facilitate implementation, but the key advantage of \harpa is that it handles unknown wave speed models. The fact that at higher rates all instances of \harpa significantly outperform all baselines points to the importance of combining wave speed adaptivity with probabilistic modeling and stochastic optimization.

\begin{figure}[!t]
    \includegraphics[width=1\textwidth]{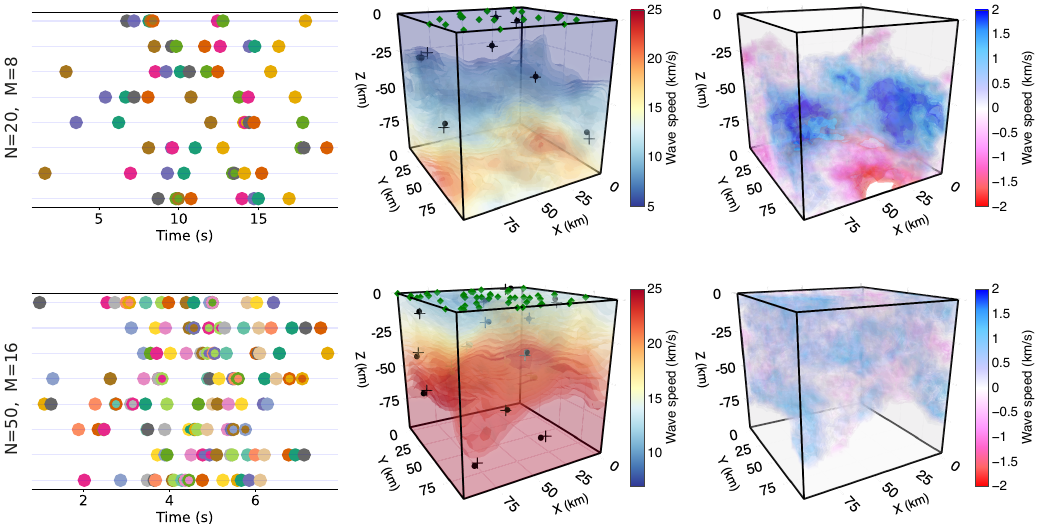}
    \caption{\harpa~association results for unknown wave speed in two instances. \textbf{Top row}: $N=20$ stations, $M=8$ sources, $\mathsf{CF}=0.93$, and  association accuracy is $91.3\%$; \textbf{Bottom row}:  $N=50$ stations, $M=16$ sources, $\mathsf{CF}=0.88$ and  association accuracy is $72.0\%$; .  \textbf{Left}: each horizontal line represents the arrivals at one station, with only 8 stations shown for illustration purposes. The color of the circle's center indicates the ground-truth association, while the edge color represents the association results of \harpa. Correct associations are indicated by consistent colors; \textbf{Middle}: background color corresponds to the wave speed; green diamonds, black crosses, and black circles represent station locations, predicted earthquake locations, and ground truth earthquake locations, respectively. \textbf{Right}: Error of the reconstruction of the wave speed model.
    }\label{fig: 8_and_16}
\end{figure}

To further stress-test \harpa's association capabilities   we test it on synthetic random field wave speed models with substantial variation, shown in Fig.~\ref{fig: 8_and_16}. The wave speed models are a superposition of a linear gradient along depth and Gaussian random fields. We simulate two settings: one with 8 sources at a frequency of 60 epm and 20 stations (\textbf{top row}) and one with 16 sources at a rate of 120 epm with 50 stations (\textbf{bottom row}). The left column shows that \harpa~accurately associates the picks even when they are completely mixed (corresponding to a confusion factor $\mathbf{CF}$ close to 1). The middle column shows that \harpa~can localize events with small error, even in depth. The right panel shows that the wave speed is also well estimated, with a maximum error below 10 percent. Notably, the results show that a greater number of both  sources and stations improve wave speed estimation.

Further details about the wave speed families used for training, specifics of the autoencoder and travel-time neural fields, as well as additional robustness studies, can be found in \textbf{SI}.

\section{Discussion }

Existing seismic phase association algorithms are optimized for low event-rate settings where the local seismic wave speed is known and/or simple. In these settings the order of wave arrivals at different stations is relatively simple~\cite{zhu2022earthquake}. The main challenge addressed by these methods is then to sieve out spurious and missing arrivals which hamper association. Our goal was to develop an entirely new paradigm for phase association which combines generative travel time neural fields and lifting arrival sequences to probability distributions.

We explore a different regime where the arrivals occur at a high rate so that the ordinal number of arrivals at a station is a priori meaningless. We also consider a challenging but realistic setting where the speed model is unknown. This regime is of great interest for earthquake monitoring but inaccessible to existing methods; see, for example, the density and mixing-up of arrivals that we successfully associate in Figs.~\ref{fig: semi} and \ref{fig: wave speed}, notwithstanding the unknown wave speed. The fact that we can still obtain accurate associations in situations of this complexity opens up new domains of application. Quoting from~\cite{adinolfi2023comprehensive}, ``a comprehensive study of microseismicity can provide a valuable description of the geological medium properties and earthquake related processes in the investigated crustal volumes, such as, for instance, the identification and geometrical characterization of active fault structures.''. 

Indeed, taming microseismicity would allow us to monitor the changes in time of the elastic material properties. One case in point is monitoring sustainable energy production facilities~\cite{grigoli2022monitoring}; another is understanding fault activity in earthquakes caused by ``unconventional hydrocarbon development'' (fracking); as Park and coauthors write~\cite{park2022basement}, ``Microearthquake activity can illuminate otherwise unknown faults, particularly when events are detected down to low magnitudes and are located precisely.''

From a high-level perspective, there are two reasons that enable \harpa to achieve this. The first is that we take an inverse problems perspective and approximately recover continuous unknowns by interpreting the observed data as a probability measure. Indeed, in situations of such complexity, it is unlikely that statistical filtering methods can work well without explicitly using some model of wave propagation. While we still only obtain (coarse) approximations of the relevant parameters, our approach paves the way toward solving the full travel time tomography problem.
We posit that it is conceivable that the resulting association and the approximate locations and wave speeds could be used as a warm start for a linearized or iterative methods~\cite{fang2020parsimonious}, or even, eventually, as a standalone approach to tomography.

A preliminary example of this is an intriguing discovery we made while applying \harpa to the Chile dataset. Letting \harpa optimize the wave speed model, we recovered a high speed heterogeneity at depths of $\geq$ 200 km. Comparing this to the known literature suggests that \harpa~recovers part of the subducting slab of the oceanic plate; the anomaly appears consistently over different wave speed parametrizations and generative models, as shown in S5 in \textbf{SI}. While a more detailed analysis of this is beyond the scope of the present paper, it lends credence to the possibility that the strategies we built in \harpa~could advance seismic tomography.

The second reason is that we leverage a body of theoretical and engineering developments from deep learning: efficient optimal transport routines, neural fields, automatic differentiation, generative models, stochastic optimization and sampling, are all essential components of \harpa.

We tested \harpa~on a suite of association tasks on real earthquake datasets and found that it outperforms SOTA models on major catalogs that have been the subject of intense recent research~\cite{zhang2019rapid, zhu2022earthquake, munchmeyer2024pyocto}. We show that when the rate increases or the wave speed model becomes more complex, SOTA algorithms become ineffective while \harpa~continues to perform accurately. The current SOTA algorithms begin to fail when the event frequency exceeds 3 events-per-minute (epm), while \harpa remains effective for much higher rates. We test complex, unknown wave speeds and scenarios with extremely dense seismic events that arrive at different stations in a completely disordered manner. We demonstrate an accurate and robust phase association in this regime. We show that \harpa~remains robust in the face of realistic wave speed distortions and noisy or missing arrival picks at the stations.

\bibliographystyle{unsrt}
\bibliography{ref}

\section*{Acknowledgments}
C.S. and I.D. are supported by were supported by the European Research Council (ERC) Starting Grant 852821—SWING. M.V.d.H. is supported by the Simons Foundation under the MATH+X program, and Department of Energy, grant DE-SC0020345.

\section*{Author contributions}
C.S, M.V.d.H. and I.D. design new models and experiments, analysed the data and wrote the paper. C.S performed the experiments.

\section*{Competing interests}
The authors declare no competing interests.

\section*{Additional information}
See \textbf{Supplementary information(SI)}

\section*{Data availability}
We utilize seismic waveform data obtained from the FDSN web service (\url{https://www.fdsn.org/webservices/}), accessed via the open-source Python API, ObsPy (\url{https://docs.obspy.org/}). The picks are then obtained through the pipeline in Seismology Benchmark collection, SeisBench (\url{https://github.com/seisbench/seisbench}). We also use the existing picks from the GaMMA project \url{https://github.com/AI4EPS/GaMMA/releases/download/test_data/demo.tar}. 
Details about the synthetic data can be found in \textbf{SI}.

\section*{Code availability}
All results in this paper are fully reproducible, and \harpa is publicly available as open-source software at \url{https://github.com/DaDaCheng/phase_association}.

\includepdf[pages=-]{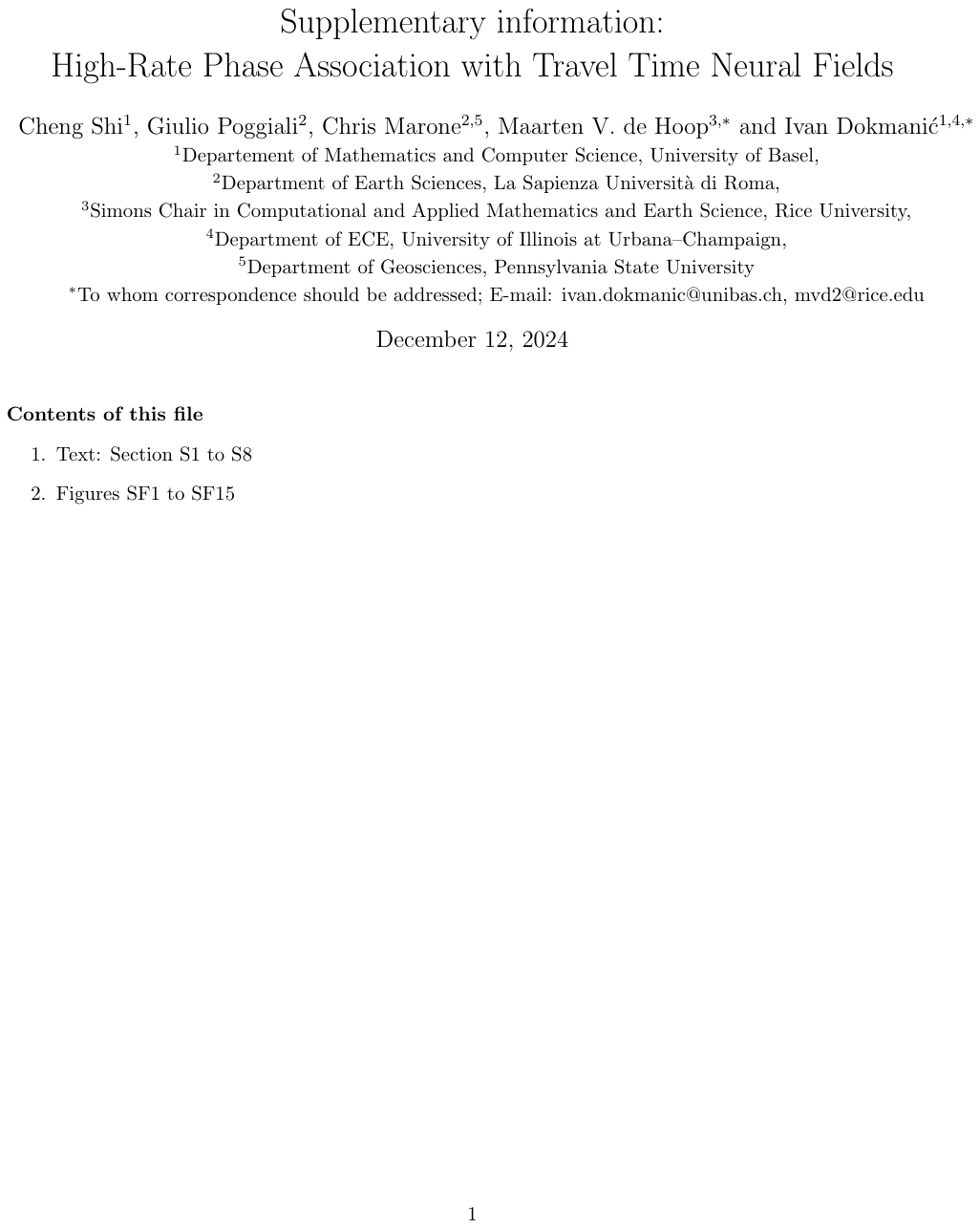}

\end{document}